\documentclass[inline,finalnew]{agujournal2019}
\usepackage{amsmath}
\usepackage{siunitx}
\usepackage{url}

\usepackage{lscape}
\usepackage{soul}

\journalname{Space Weather}

\begin{document}


\title{Parameter Distributions for the Drag-Based Modeling of CME Propagation}


\correspondingauthor{Gianluca Napoletano}{gianluca.napoletano@roma2.infn.it} 

\authors{
	Gianluca Napoletano\affil{1} 
,
    Raffaello Foldes\affil{2,3}
,
    Enrico Camporeale\affil{5,6}
,
    Giancarlo de~Gasperis\affil{1,4}
,
    Luca Giovannelli\affil{1}
,
    Evangelos Paouris\affil{7,8}
,
    Ermanno Pietropaolo\affil{2}
,
    Jannis Teunissen\affil{9}
,
    Ajay Kumar Tiwari\affil{9}
,
    Dario Del~Moro\affil{1,4}
} 
    
\affiliation{1}{
Dipartimento di Fisica, Universit\`a degli studi di Roma ``Tor Vergata'',  Viale della Ricerca Scientifica 1, I-00133 Roma, Italy}
\affiliation{2}{
Dipartimento di Scienze Fisiche e Chimiche, Universit\`a degli studi dell'Aquila, Via Vetoio 42, I-67100 Coppito AQ, Italy}
\affiliation{3}{
Laboratoire de Mécanique des Fluides et d’Acoustique, CNRS, École Centrale de Lyon, Université Claude Bernard Lyon 1, INSA de Lyon, F-69134 Écully, France}
\affiliation{4}{
INFN sezione di Roma2, Viale della Ricerca Scientifica 1, I-00133 Roma, Italy}
\affiliation{5}{
CIRES, University of Colorado, Boulder, CO, USA}
\affiliation{6}{
{NOAA Space Weather Prediction Center, Boulder, CO, USA}\\
\affiliation{7}{
Institute for Astronomy, Astrophysics, Space Applications \& Remote Sensing of the National Observatory of Athens, Penteli, Greece}
\affiliation{8}{
Faculty of Physics, National and Kapodistrian University of Athens, Athens, Greece}
\affiliation{9}{
Centrum Wiskunde $\&$ Informatica (CWI), Amsterdam, The Netherlands}
}

\begin{keypoints}
\item A new CME-ICME database is created from observations from multiple sources.
\item  Statistical distributions for the model parameters are determined using the database and the DBM equations, taking uncertainties into account.
\item The PDBM is updated with the new distributions and validated with a different CME database.
\end{keypoints}


 \begin{abstract}
In recent years, ensemble modeling has been widely employed in space weather to estimate uncertainties in forecasts. We here focus on the ensemble modeling of CME arrival times and arrival velocities using a drag-based model, which is well-suited for this purpose due to its simplicity and low computational cost. Although ensemble techniques have previously been applied to the drag-based model, it is still not clear how to best determine distributions for its input parameters, namely the drag parameter and the solar wind speed. The aim of this work is to evaluate statistical distributions for these model parameters starting from a list of past CME-ICME events . We employ LASCO coronagraph observations to measure initial CME position and speed, and in situ data to associate them with an arrival date and arrival speed. For each event we ran a statistical procedure to invert the model equations, producing parameters distributions as output.
Our results indicate that the distributions employed in previous works were appropriately selected, even though they were based on restricted samples and heuristic considerations. On the other hand, possible refinements to the current method are also identified, such as the dependence of the drag parameter distribution on the CME being accelerated or decelerated by the solar wind, which deserve further investigation.
\end{abstract}

\section*{Plain Language Summary}
Coronal Mass Ejections, consisting of huge expulsions of plasma and magnetic field from the solar corona, are important for space weather. Among several forecasting techniques, the drag-based model, which describes CME propagation in interplanetary space, is widely used to compute CME transit time and impact speed, by describing the CME propagation as that of a solid body moving in an external fluid.
 In recent years, this model has been improved via a new approach in which statistical distributions of the input quantities are introduced to evaluate uncertainties of the resulting forecasts. Unfortunately, such distributions for the model parameters are still not very well known from experimental observations and it is hard to obtain them from theoretical models. In this work, we built an empirical method to evaluate such statistical distributions using a list of past CME-ICME events. New findings emerged from this analysis, such as a dependence of the drag parameter on the ICME being accelerated or decelerated, deserve further investigation.

\section{Introduction}
Coronal Mass Ejections \cite<CMEs ->{chen2011coronal, chen2017physics, webb2012coronal} are expulsions of plasma and magnetic field from the Sun.
Their interplanetary counterparts, ICMEs  \cite{kilpua2017coronal} are among the main drivers of the space weather \cite<SWx ->{schwenn2006space, pulkkinen2007space} with impact on the whole heliosphere, and they are responsible for the strongest variations in the near-Earth solar~wind conditions \cite<e.g.,>{tsurutani1988origin, schwenn2005association, buzulukova2017extreme}.
These variations trigger a number of effects on space-borne and ground-based technologies, either directly or via major geomagnetic storms.

The accurate prediction of the arrival and characteristics of ICMEs at Earth, and more recently elsewhere in the heliosphere, is a necessity to minimize the impact on the existing and future assets, and has always been a primary goal of the SWx forecasting \cite<see e.g.,>{daglis2001space,schrijver2010heliophysics, berrilli2017swerto, iwai2019development, veettil2019ionosphere}.
Note that in this work the terms CME and ICME refer to the plasma and magnetic field structure expelled from the Sun, without the shock that precedes it.

Forecasting the Time of Arrival (ToA) and Speed of Arrival (SoA) of an ICME more than an hour ahead is nevertheless a complicated task, since it requires to understand the propagation of a poorly determined plasma and magnetic field structure into an essentially undetermined interplanetary environment. 
A recent review \cite{vourlidas2019predicting} assessed the current state of ICME ToA and SoA forecasting algorithms by surveying the recent literature.
While there is quite a scatter in the results and perhaps a bias correlated to the sample size used, the authors report that the most recent forecasting methods have a mean absolute error (MAE) close to 10 hours, which is similar in the case of empirical, simplified physics or full MHD models \cite<as ENLIL or EUHFORIA>{odstrcil2003modeling, pomoell2018euhforia}, or machine learning based \cite<see>[for a list of such approaches]{camporeale2019challenge}.
\citeA{vourlidas2019predicting} conclude that a number of factors (physical, observational and modeling) are limiting the performances of all approaches, and that the ToA accuracy in particular is limited by the quality of the currently available data. 
These complications arise from the difficulty to evaluate remotely the properties of the CME at launch with the present-day instrumentation and from the impossibility to properly characterize the status of the inner heliosphere.
Therefore, in order for the forecast to be useful, it should cope with this lack of information and provide an estimate of its intrinsic uncertainty \cite{owens2020computationally}.
This can be achieved by empirical methods through the use of statistical relationships established between past CME measured parameters and ICME characteristics \cite{gopalswamy2001predicting, kilpua2012estimating},     
or can be achieved by numerical MHD-based models by using ensembles of runs to model the same ICME with different initial conditions that represent the inherent uncertainties \cite{emmons2013ensemble, mays2015ensemble, cash2015ensemble}.

The main shortcoming of the statistical approach is that it ``treats all events the same and neglects the contextual information and knowledge that certain situations are inherently more predictable than others'' \cite{owens2020computationally}.
The other approach (i.e. treat every CME as a different case) has to cope with a relatively large parameter space to explore, and the relatively long time needed for the computation of each simulation run (of the order of tens of minutes on high-performance systems).

A possible solution is to adopt simplified, kinematic models, for instance, assuming a simplified solar wind propagation, a simple CME geometry, and a hydrodynamic-like ICME-solar wind interaction.
Among these models, the drag-based model \cite<DBM - >{cargill2004aerodynamic, vrsnak2013propagation}, is among the most used and it can be run in large ensembles during a few seconds in an average laptop
\cite{amerstorfer2018ensemble,dumbovic2018drag,kay2018effects,napoletano2018probabilistic}.
The DBM requires CME properties as input: the CME launch time, initial speed, direction and angular width. 

To model the interaction of the ICME with the background solar wind the DBM needs only the solar wind speed and the value of the drag parameter $\gamma$, which determines the interaction between the solar wind and the CME.
The quantities used to describe the CME are retrieved from observations that have associated measurement errors. They can be used in ensemble models assuming Gaussian probability distribution functions (PDFs). 
The solar wind speed and the drag parameter are instead drawn from \emph{a-priori} PDFs, modeled from empirical PDFs built from past data sets of CME and associated ICME characteristics measured at liftoff and at Lagrange point L1, respectively.
The outputs of the DBM ensemble model are the PDFs of ToA and SoA at a target location.
From these, we can estimate the most probable ToA and SoA, and their associated prediction uncertainties \cite<e.g.,>{del2019forecasting, piersanti2020sun}.\\
In recent years there has been several interesting development of DBM based tools to forecast ICME arrival time and speed. 
When compared against other forecasting methods as in \cite{vourlidas2019predicting}, their performance is comparable and sometimes even better than MHD based methods \cite{vrsnak2014heliospheric}. 
Augmenting the results presented in Table 1 of \cite{vourlidas2019predicting} with Table 2 of \cite{dumbovic2021drag}, we can estimate that the typical MAE of the DBM methods is of the order of 10h and the typical error on the SoA is around 50 km/s.
It deserves to be stressed that the performance of each model has been computed on a different ICME dataset, thus limiting the value of this comparison. A community effort to create a common benchmark is underway \cite<e.g.,>{verbeke2019benchmarking}, but it is not yet a widespread standard.\\
Recently, \citeA{kay2020identifying} used the DBM ensemble approach to answer an implicit question in \citeA{vourlidas2019predicting}: ``How much should we improve our knowledge of the parameters to improve the ToA predictions beyond the present limit MAE of about 10 hr''?
Their conclusion is that the most critical parameter is the CME speed at liftoff, and efforts should be spent to achieve a better and more homogeneous determination of the CME speed via coronagraphics imaging or other means.  
\citeA{kay2020identifying} also explored the sensitivity of the ToA vs the $\gamma$ parameter, following \citeA{dumbovic2018drag}, who postulated a symmetric distribution of values ($\gamma = 0.1\pm 0.05 \times 10^{-7}\si{km^{-1}}$ ).
However, \citeA{napoletano2018probabilistic} (hereafter Paper~I) built an empirical $\gamma$ PDF which is asymmetrical and not compatible with a Gaussian shape.
Since the $\gamma$ parameter incorporates much of the physics of the ICME-wind interaction and its precise value is poorly understood \cite{kay2020identifying}, we think it deserves further investigation.  
In particular, the empirical $\gamma$ PDF in Paper~I has been built from a limited database of ICME, therefore it is the quantity that most of all needs a more robust definition. A better assessment of the values of $\gamma$ may be obtained, in principle, with detailed knowledge of the ICME kinematics. In order to evaluate the DBM input parameters, \cite{Zic2015} proposed to employ a least-squares fitting procedure to the ICME interplanetary tracking derived from STEREO coronagraphic and heliospheric image observations. In addition, \cite{Rollett2016} also developed such an approach and employed an improved model for the geometrical shape of the CME front, showing that the extrapolation of the CME dynamics based on real-time tracking can further reduce the mean error of the predicted arrival time to $6.4\pm5.3 \si{h}$ and impact speed to $16 \pm53 \si{km/s}$.
Although promising, such an approach is currently not feasible for real-time forecasting, as it requires a dedicated heliospheric observatory for real-time ICME tracking. 
Therefore model parameters have to be constrained by two single observations (at liftoff and Earth), and a better assessment of these parameters depends on the availability of more data and an improved database.
The aim of this paper is to use a large number of ICMEs to build a new empirical PDF for the model parameters and to find a suitable functional form to model it.

The paper is structured as follows.
In Section~\ref{Data} we describe the data and the methods employed to build the ICME database used to obtain the new $\gamma$ PDF.
Section~\ref{Methods} contains the technical description of the methods used to retrieve the $\gamma$ PDF.
Section~\ref{Results} contains the actual results and a validation of those results against the ICME list from \citeA{paouris2017interplanetary}.
In Section~\ref{Disc+Conc} we discuss the results and provide a synopsis of the findings of this paper.
The ICME catalog built for this analysis, together with a tool for the data set visualization and the software employed for the ensemble simulation through the probabilistic drag based model, is available from \url{https://zenodo.org/record/5517629} \cite{Napoletano2021}.

\newpage
\section{Data}
\label{Data}
As mentioned above, the DBM needs input values to output a ToA and a SoA at a target location.
For the purpose of this work, we compare observed ToA and SoA at Earth (or L1) position against those computed by the probabilistic drag-based model (PDBM) presented in Paper~I.
We therefore need a database associating Earth ToA and SoA of an ICME to the kinematic characteristics of the corresponding CME.
In particular, we need measures of the position $r_0$ and the speed $v_0$ at time $t_0$ of the CME front, and the solar wind speed (see~\ref{AppendixA3}).\\

Also, to obtain a homogeneous database for a consistent approach, we re-computed some of the CME kinematics properties (CME leading edge initial initial speed and acceleration) with a standardized method.
Last, to have an assessment of the method as close as possible to the actual operation performances, we made use of the same methods and algorithms presently implemented in the real-time ICME ToA forecast running at \url{http://spaceweather.roma2.infn.it}. 
These selection criteria and methods are described in the following sub-sections.

\subsection{Sources for compiling the employed ICMEs database}
Our analysis uses a database connecting the kinematic parameters of the CME at launch time and the information about the arrival time and speed of the related ICME. 
The databases already available are not suited for our analysis, but we can merge and complement the information from three different sources available online to create a new database for this purpose.

We start from the list of near-Earth interplanetary coronal mass ejections (ICMEs) compiled by \citeA[hereafter R\&C, 2010]{richardson2010near}, who maintain the catalog using data from the OMNI database (Goddard Space Flight Center, GSFC, \url{http://spdf.gsfc.nasa.gov/}).
From about 500 events from May 1996 to present in this catalog, we consider only those events that have information about the CME liftoff time, selecting 247 events.
This onset time  (see \url{http://www.srl.caltech.edu/ACE/ASC/DATA/level3/icmetable2.htm} for specifications) refers to the most-likely association between the ICME and the corresponding CME observed by LASCO coronagraphs (C2 and C3) aboard SOHO spacecraft (in a few cases, STEREO spacecraft observations are used).

We make use also of information about the CME liftoff retrieved from the SOHO LASCO CME catalog (\url{https://cdaw.gsfc.nasa.gov/CME_list/}) maintained at the CDAW Data Center. 
Lastly, information about features and events on the Sun that may be associated to the CME, such as Filaments, Flares, Active Regions, and Coronal Holes are retrieved from the Heliophysics Event Knowledgebase \cite[HEK]{hurlburt2010heliophysics, martens2012computer}: a queryable repository of feature and event information about the Sun.

\subsection{Database creation}
We associate each of the remaining 247 ICME in the R\&C list with an entry in the CDAW SOHO/LASCO CME catalog having the same onset times, $t_{0_\text{R\&C}}=t_{0_\text{LASCO}}$, by using a dedicated Python routine. 
The SOHO/LASCO catalog provides various kinematic properties associated to the onset time $t_{0_\text{LASCO}}$ of the CMEs (i.e. \emph{position angle}, \emph{angular width}, \emph{Mass}, see Tab.~\ref{DataTab}).
Among those properties, there is also an estimate for the CME lift-off speed, but unfortunately, it is relative to the fastest part of the CME front and has no associated error.
Therefore, it can not be employed to define a PDF and is not suitable for our purposes. Nevertheless, with this association, we can retrieve from the SOHO/LASCO website the original height versus time measurements of the CME front (in solar radii units, $R_\text{Sun}$), computed from LASCO~C2/C3 coronagraph images.
These values represent the position of the CME front on the plane-of-sky (hereafter POS), and need to be de-projected to obtain the true radial distance and speed of the CME.
The equation to de-project the position $r_{0_\text{POS}}$ into $r_0$ is from \cite{gopalswamy2010deprojection}.
It implies a model for the CME shape \cite<i.e. the CME front expansion is considered completely radial, as in Model~A of Figure~9 in >[]{schwenn:hal-00317653} and requires the CME angular width and the location of the CME source on the solar disk.

\begin{figure}[p!]
\includegraphics[width=\textwidth]{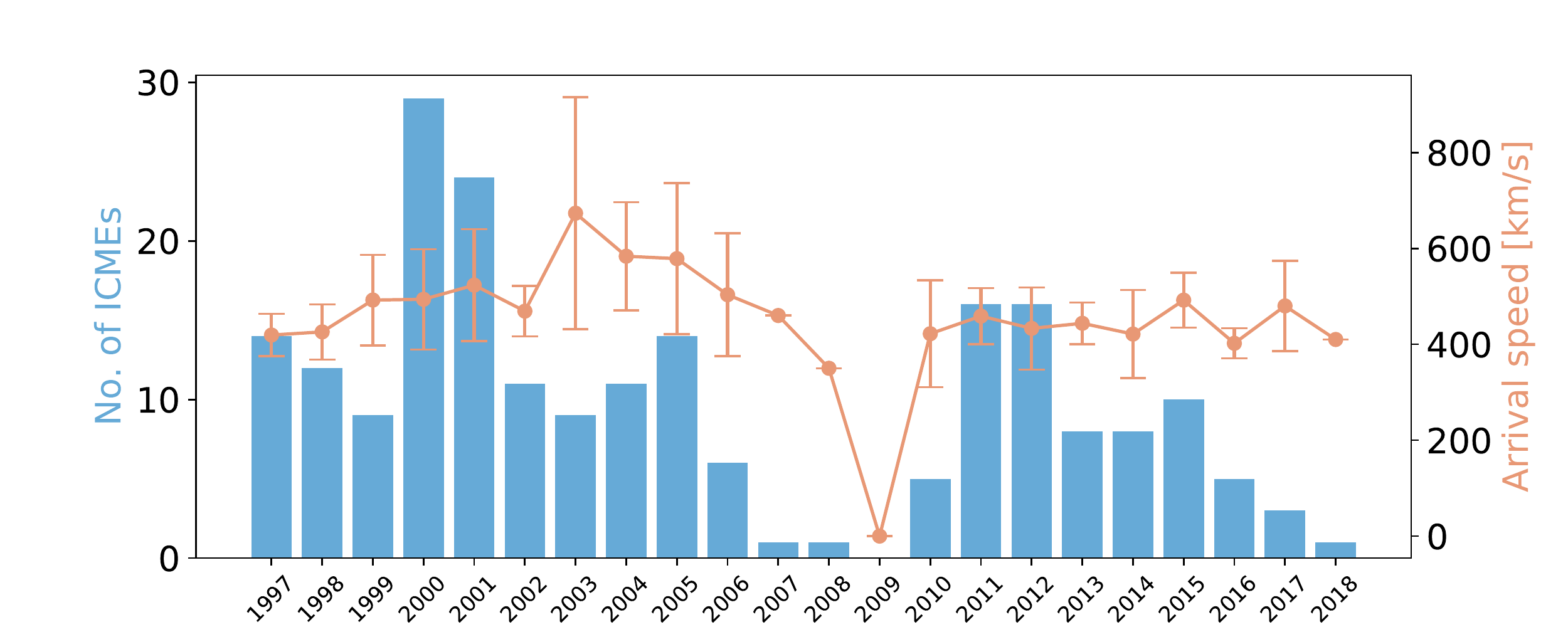}
\caption{
Distribution of the selected ICMEs in our data set during the past 20 years. The orange line represents the average SoA of ICMEs for each year, the error bar represents the relative standard deviation.
}
\label{CME_per_year}
\end{figure}

\begin{figure}[p!]
\makebox[\textwidth][c]{\includegraphics[width=\textwidth]{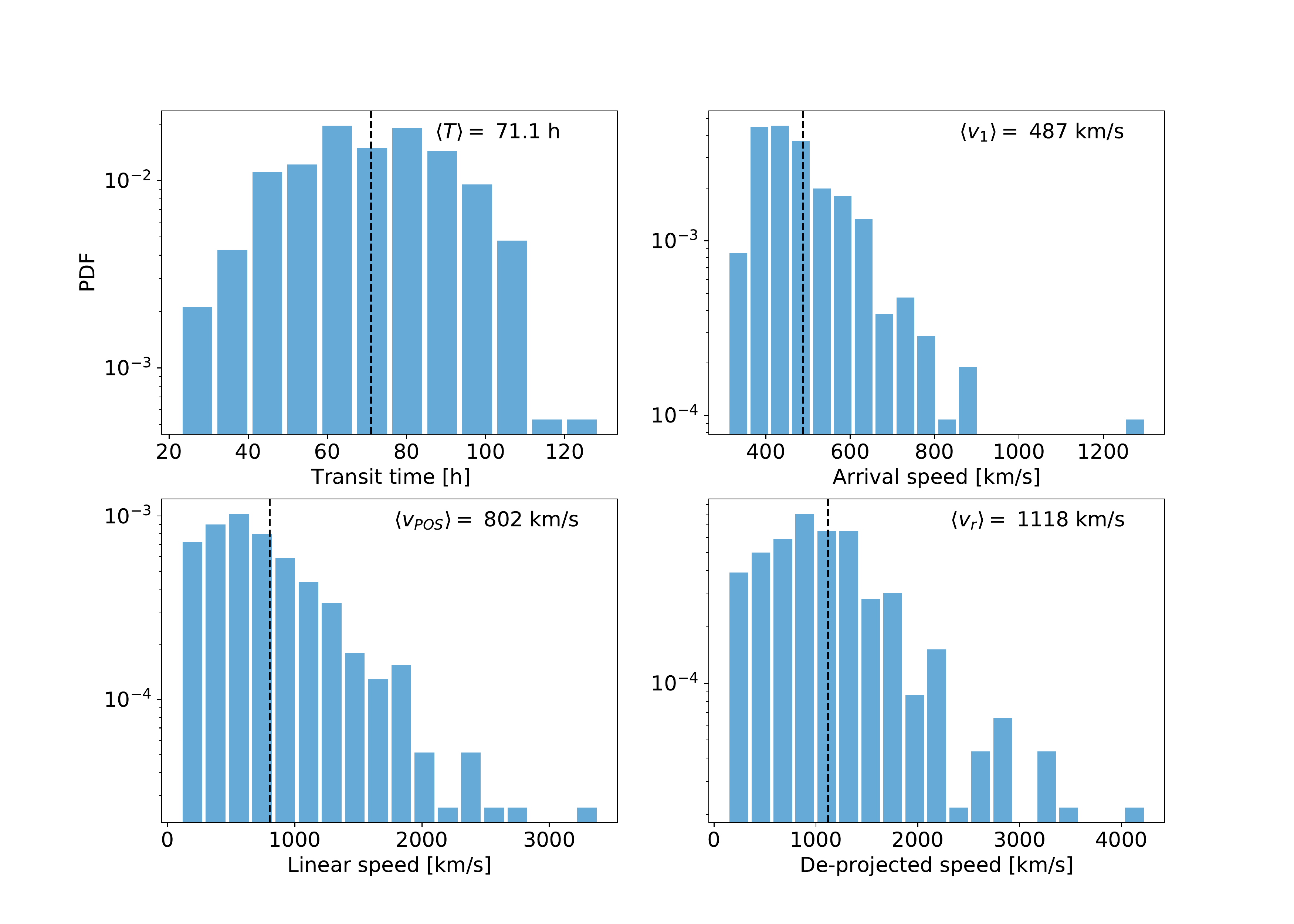}}
\caption{(From left to right) Probability distribution of the ICME ToA, SoA, linear plane-of-sky liftoff speed and de-projected liftoff speed obtained with from whole ICME data set. In every panel, the vertical dashed line (black) is the average value, also reported within each panel.}
\label{image2}
\end{figure}

The details about the algorithms to associate a CME to its source on the solar disk and to compute its de-projected speed at $r_0=20R_\text{Sun}$ and the associated error are described in the ~\ref{AppendixA}.\\
\noindent
An interesting new geometrical technique for de-projecting the CME speed has been very recently published \cite{paouris2021b}. We foresee the possibility to utilize such a method in a future work.
After these procedures our data set was reduced to 214 CME-ICME pairs. In Figure~\ref{CME_per_year} we report the histogram of the yearly number of CME comprising this data set, since it spans about two solar cycles (23 and 24). As it is known \cite<e.g.,>[]{webb2012coronal, 2019SSRv..215...39L}, the number of CMEs depends on the solar cycle phase and the total number of CMEs per cycle is clearly different  for thecycle 23 and 24. In Figure~\ref{image2} we show a summary of some quantities reported in Tab.~\ref{DataTab}, and in particular we focus on those variables that will be used for the inversion procedure (see Sec.~\ref{sec:inv_procedure}).

%
\clearpage
\begin{landscape}
\begin{table}[p]
\centering
\caption{Column description of the ICMEs data set created in this work.} 
\begin{tabular}{c l l p{10.7cm} c}
\hline
  & Name  & Keyword & Description & Source\\
\hline
  1 & \emph{LASCO start}  & LASCO\_Start & First CME appearance in LASCO C2/C3 coronographs & LASCO/CDAW\\
  2 & \emph{Start date}  & Start\_Date & Start time of CME extrapolated at 20 $R_{\odot}$ & This work\\
  3 & \emph{Arrival date}  & Arrival\_Date & Estimated arrival time of the ICME based primarily on plasma and magnetic field observations & R\&C\\
  4 & \emph{Plasma event dur.}  & PE\_duration & End of the ICME plasma signatures after col. 3 is recorded & R\&C\\
  5 & \emph{Arrival speed}  & Arrival\_v & ($km/s$) ICME arrival speed measured at L1 ($\sim$ 1AU) & R\&C\\
  6 & \emph{Transit time}  & Transit\_time & ($hrs.$) Computed between col. 3 and col. 1 & This work\\
  7 & \emph{Trans. time error}  & Transit\_time\_err & ($hrs.$) Error associated to the extrapolated start date (col. 3) of a CME & This work\\
  8 & \emph{LASCO date}  & LASCO\_Date & Most likely CME associated with the ICME observed by LASCO & LASCO/CDAW\\
  9 & \emph{LASCO speed}  & LASCO\_v & ($km/s$) Max. plane-of-sky (POS) CME speed along the angular width  & LASCO/CDAW\\
  10 & \emph{Position angle}  & LASCO\_pa & ($deg.$) Counterclockwise (from solar North) angle of appearance into coronographs & LASCO/CDAW\\
  11 & \emph{Angular width}  & LASCO\_da & ($deg.$) Angular expansion of CME into coronographs & LASCO/CDAW\\
  12 & \emph{Halo}  & LASCO\_halo & If col. 15 is $>$ \ang{270} then 'FH' (full halo), if $>$ \ang{180} 'HH' (half halo), if $>$ \ang{90} 'PH' (partial halo), otherwise 'N'. & LASCO/CDAW\\
  13 & \emph{De-proj. speed}  & v\_r & ($km/s$) De-projected CME speed (from 9, see Appendix \ref{AppendixA1}) & This work \\
  14 & \emph{De-proj. speed error}  & v\_r\_err & ($km/s$) Uncertainty of CME initial speed (col. 13) & This work\\
  15 & \emph{Theta source} & Theta & ($arcsec$) Longitude of the most likely source of CME& This work\\
  16 & \emph{Phi source} & Phi & ($arcsec$) Co-latitude of the most likely source of CME & This work\\
  17 & \emph{Source pos. error} & POS\_source\_err & ($deg.$) Uncertainty of the most likely CME source & This work\\
  18 & \emph{POS source angle} & POS\_source & ($deg.$) Principal angle of the most likely CME source & This work\\ 
  19 & \emph{Relative width} & rel\_wid & ($rad.$) De-projected width of CME & This work\\
  20 & \emph{Mass}  & Mass & ($g$) Estimated CME Mass (if provided) & LASCO/CDAW\\
  21 & \emph{Solar wind type}  & SW\_type & Solar wind (slow, S, or fast, F) interacting with the ICME & This work\\
  22 & \emph{Bz}  & Bz &($nT$) $z$-component of magnetic field at L1 and CME arrival time (col. 3) & R\&C\\
  23 & \emph{Dst}  & DST & Geomagnetic Dst index recorded at CME arrival (col. 3) & R\&C\\
  24 & \emph{Stat. de-proj. speed}  & v\_r\_stat & ($km/s$) Statistical de-projected CME speed, i.e. v\_r\_stat=LASCO\_v*1.027+41.5 & Paouris et al.\\
  25 & \emph{Acceleration} & Accel & ($m/s^2$) Residual acceleration at last CME observation & This work\\
  26 & \emph{Analytic sol. wind} & Analytic\_w & ($km/s$) solar wind from DBM exact inversion & This work\\
  27 & \emph{Analytic gamma} & Analytic\_gamma & ($km^{-1}$) drag parameter, $\gamma$, from DBM exact inversion & This work\\
\hline
\end{tabular}
\label{DataTab}
\end{table}%
\end{landscape}

\newpage
\section{Methods}
\label{Methods}
\subsection{The Drag-Based Model}
We employ the drag-based model \cite<DBM>[]{cargill2004aerodynamic, vrsnak2013propagation} to forecast arrival time and impact speed of a CME at Earth.
This model assumes that, from a certain distance from the Sun, the CME dynamics is governed only by its interaction with the ambient solar wind.
By employing a fluid dynamic analogy, it is assumed that the force depends on the square of the relative velocity of the CME to the ambient solar wind flow, so that the equation for the CME radial acceleration reads:
\begin{equation}
\label{DBM1}
    a =  -\gamma(r) \left[v-w(r) \right]\left|v-w(r)\right|
\end{equation}
where $\gamma(r)$ is the so-called drag parameter, representing the interaction efficiency between the CME and the solar wind, $w(r)$ is the solar wind speed, and $r$ is the distance from the Sun. 
A reasonable approximation beyond 20 solar radii is that of constant $\gamma$ and $w$ throughout the whole ICME propagation \cite{cargill2004aerodynamic, vrsnak2013propagation}. We point out that this is a relevant assumption, as in reality the ambient solar wind speed and the mechanisms of interaction between the solar wind and the CME structure are not constant. We refer the reader e.g. to \cite{Temmer_2012}, \cite{Rollett_2014}, \cite{Zic2015} for studies investigating CMEs evolving in different drag regimes and variable solar wind speed, or \cite{piersanti2020sun} for an example of the probabilistic approach to the drag-based model applied with a variable solar wind.
Under such assumptions, equation~(\ref{DBM1}) can be solved analytically \cite{vrsnak2013propagation} for the heliospheric distance and the ICME speed as a functions of time:
\begin{equation}
\label{DBMr}
    r(t) = \pm \frac{1}{\gamma} \ln{[1\pm\gamma(v_0-w)t]}+wt+r_0,
\end{equation}
\begin{equation}
\label{DBMv}
    v(t) =  \frac{v_0-w}{1\pm\gamma(v_0-w)t}+w,
\end{equation}
where the choice between $\pm$ depends on the sign of $v_0 - w$ : the + sign is taken for accelerated CMEs ($v_0 - w < 0$), while the - sign holds for decelerated ones ($v_0 - w > 0$).
Given the initial conditions $r_0, v_0$ and model parameters $\gamma$ and $w$, these equations can be employed to compute the ToA and the SoA at a target located at a chosen heliocentric distance.

\subsection{The inversion procedure}\label{sec:inv_procedure}
The DBM equations (\ref{DBMr}) and (\ref{DBMv}) can be  used to forecast the travel time $T$ and the impact speed $v_1$ of an ICME at a given position $r_1$. 
Conversely, if $T$ and $v_1$ are known, these equations can be inverted leaving $\gamma$ and $w$ as unknown values, as in  \cite{vrsnak2013propagation}: 
\begin{equation}
\label{implicit1}
r_1 = r_0 + wT+\frac{(v_0-w)(v_1-w)\,T}{(v_0-v_1)}\ln{\left[ 1+\left(\frac{v_0-v_1}{v_1-w}\right)\right]}     
\end{equation}
\begin{equation}
\label{implicit2}
    \pm \gamma = \frac{(v_0-v_1)}{(v_1-w)(v_0-w)T}
\end{equation}
The first equation is implicit in $w$, but can be solved numerically and its solution is then employed to compute $\gamma$ through the second one.
Therefore, for a given $T, v_0, v_1$, we have $\gamma=f(w)$.
This dependence can be seen in the joint PDFs presented below, where for each CME there is a ridge of joint $w,\gamma$ values.

In principle, using the entries of the database presented in Section~\ref{Data}, we could compute the $\gamma$ and $w$ values for every ICME and add them to the database.
In practice, due to the intrinsic errors of the $r_0, r_1, v_0, v_1$ and $T$ parameters, or due to the fact that sometimes the DBM model simply does not accurately represent the ICME motion (e.g. in the case when $w=\text{constant}$ is not a viable approximation) equations (\ref{implicit1}) and (\ref{implicit2}) have no  solution for 75 out of the 214 events of our database. 
We therefore used an inversion method that takes into consideration the experimental uncertainty of the input quantities during the inversion procedure.
The two parameters $r_0$ and $r_1$ have no associated errors, since we set them at 20 and 215 solar radii, respectively. 
For $v_0$, $v_1$ and $T$, we modeled the PDFs with normal distributions centered at the relative measured or estimated values and with standard deviations corresponding to their uncertainties.
For $v_0$, the mean is defined by the CME de-projected speed, and the standard deviation $\sigma$ by its associated error, rows 13 and 14 in Tab.~\ref{DataTab} respectively.
For $v_1$, the mean is defined by the ICME measured arrival speed (row 5 in Tab.~\ref{DataTab}), and $\sigma$ by an assumed measurement error of $10\%$.
For $T$, the mean is defined by the  difference between the measured arrival time at Earth and the estimated passage at $20R_{\text{Sun}}$ row 6 in Tab.~\ref{DataTab}, and $\sigma$ by considering the error on the liftoff time obtained by the de-projection procedure (row 7 in Tab.~\ref{DataTab}).

From the average values and their PDFs, the inversion method generates $N$ random samples of $[r_0, r_1, v_0, v_1, T]$ per ICME and feeds those to equations (\ref{implicit1}) and (\ref{implicit2}).
We run this inversion procedure with $N=5000$ for each of the 214 events in the ICME database.
In about 50\% of all the generated cases the solutions $\gamma$ and $w$ do not exist due to incompatibility of the randomly generated input values,i.e., such values do not allow for a solution of the implicit equation (\ref{implicit1}).

Also, we take into consideration for the following analysis only those solutions where $10^{-8} \text{km}^{-1} < \gamma < 10^{-6} \text{km}^{-1}$.
We choose this range of magnitudes considering Eq. 2 of \citeA{vrsnak2013propagation} and following the same order of magnitude reasoning therein, which poses a limit on the realistic values of $\gamma$.\\
\newpage
\section{Results}
\label{Results}
\subsection{Inversion method results}
\begin{figure}[p!]
\centering
\includegraphics[width=\textwidth]{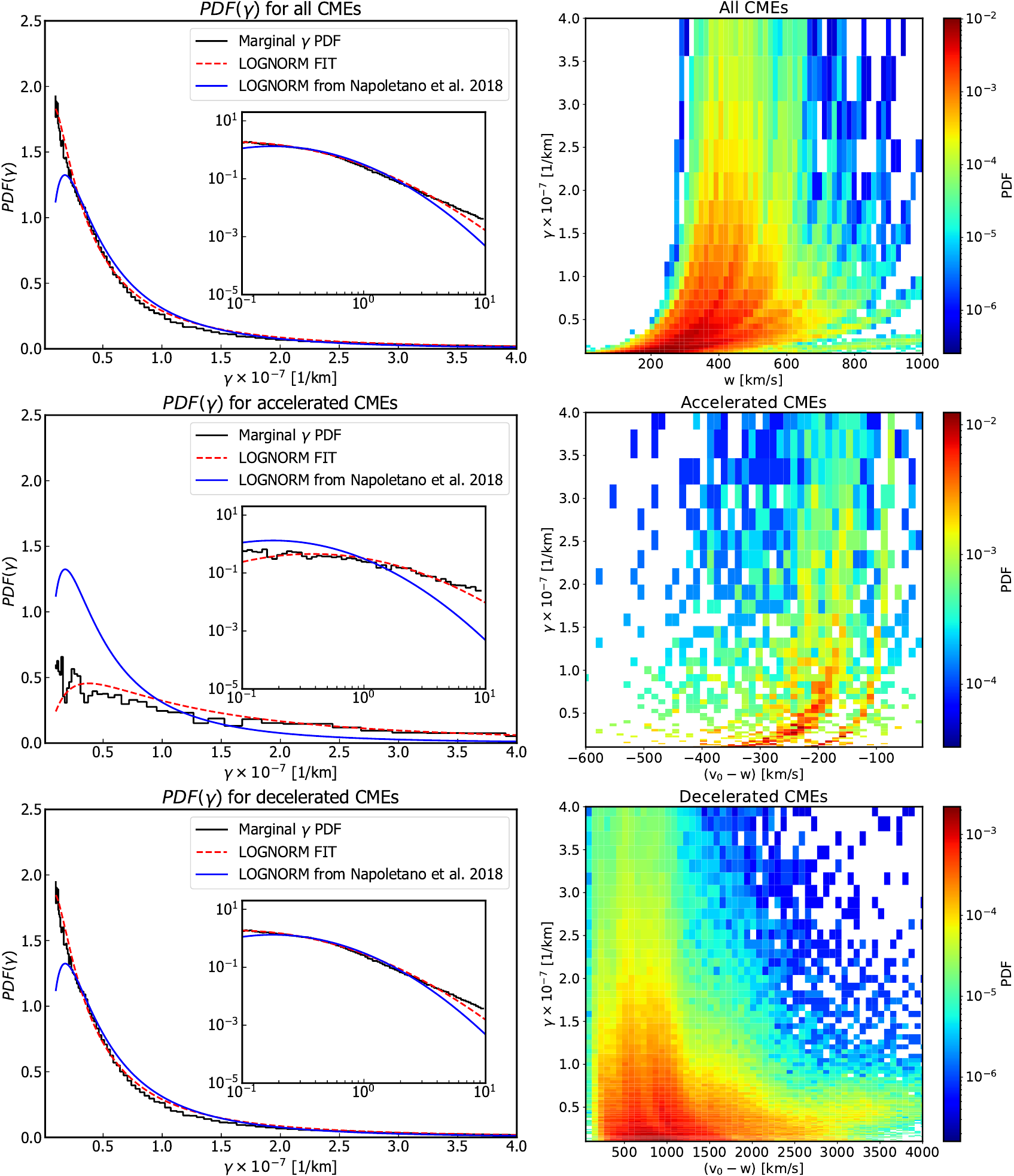}
\caption{PDFs for $\gamma$ (left column) and the joint distributions (right column) obtained from inversion procedure using the ICMEs in the database as input.
Upper panel: $\gamma$ vs $w$ for the whole dataset (519,857 values).
Middle panel:  $\gamma$ vs $\Delta v$ for the accelerated ICMEs (18,416 values).
Lower panel:  $\gamma$ vs $\Delta v$ for the decelerated ICMEs (501,441 values).
All the PDFs were fitted with a lognormal function (red dashed). The lognormal function (blue line) used in Paper~I is also plotted for comparison.
 In each panel, the inset shows the same plot on a log-log scale for an extended range.}
\label{image3}
\end{figure}
The inversion procedure was successful for 210 out of 214 events, thus providing a statistical distribution for the solar wind $w$ and drag parameter $\gamma$. The whole sample consists of 519,857 inversions.
In the upper-left panel of Figure~\ref{image3}, we show the joint distribution $\gamma-w$.
As already stated in Par.~\ref{sec:inv_procedure}, in the joint distribution we can still identify a few $w-\gamma$ ridges generated by single CMEs, but the plane is populated enough to extract the properties of these PDFs and compare them with the PDFs used in Paper~I.
Also, as a consequence of the random extraction of the initial speed and solar wind, we can draw two more joint PDFs, separating the accelerated $\Delta v = v_0 - w < 0$ (18,416 extractions) and decelerated $\Delta v > 0$ (501,441 extractions) ICMEs. 
These two joint PDFs  $\gamma-\Delta v$ are shown respectively in the central-left and lower-left panels of Figure~\ref{image3}.
%
%

\subsection{PDFs for the solar wind speed and drag parameter}
\label{sub:PDFs}

From the joint distributions shown in Figure~\ref{image3}, we can extract the marginal distributions for the drag-parameter, $\gamma$.
We compare this empirical PDF with the reference lognormal function used in Paper~I to model the $\gamma$ PDF.
The plot in the upper panel of Figure~\ref{image3} shows that the histogram retrieved from the whole dataset has a shape which is not compatible with the reference PDF (in blue).
The fit of the histogram with a lognormal function (red dashed) retrieves $\mu=-0.83$ and $\sigma=1.26$, against the $\mu=-0.70$ and $\sigma=1.01$ of the reference PDF. 
It is worth to note that we tried fitting to the PDFs other function types, in particular exponentials, power laws, and truncated power laws, but none of those retrieved a better fit than the lognormal.\\
As above, we can divide our dataset in accelerated and decelerated ICMEs (central left and lower left panels of Figure~\ref{image3}).
The two histograms look quite different and are fitted by lognormal functions (red dashed) with significantly different parameters.
%
In particular, from the fit to the histogram from the decelerated ICMEs, we retrieve $\mu=-0.85$ and $\sigma=1.25$. 
Since this histogram contains more than 98\% of the total samples, its parameters are close to those obtained from the whole dataset, but also more compatible with the values of Paper~I.\\
From the fit to the histogram from the accelerated ICMEs, we retrieve $\mu=0.40$ and $\sigma=1.18$.  
The accelerated CMEs are represented by a lognormal distribution with a significantly higher mean value.\\
From these results, we define new $\gamma$ parameter distributions functions for accelerated and decelerated cases.
\begin{figure}[p!]
\centering
\includegraphics[width=\textwidth]{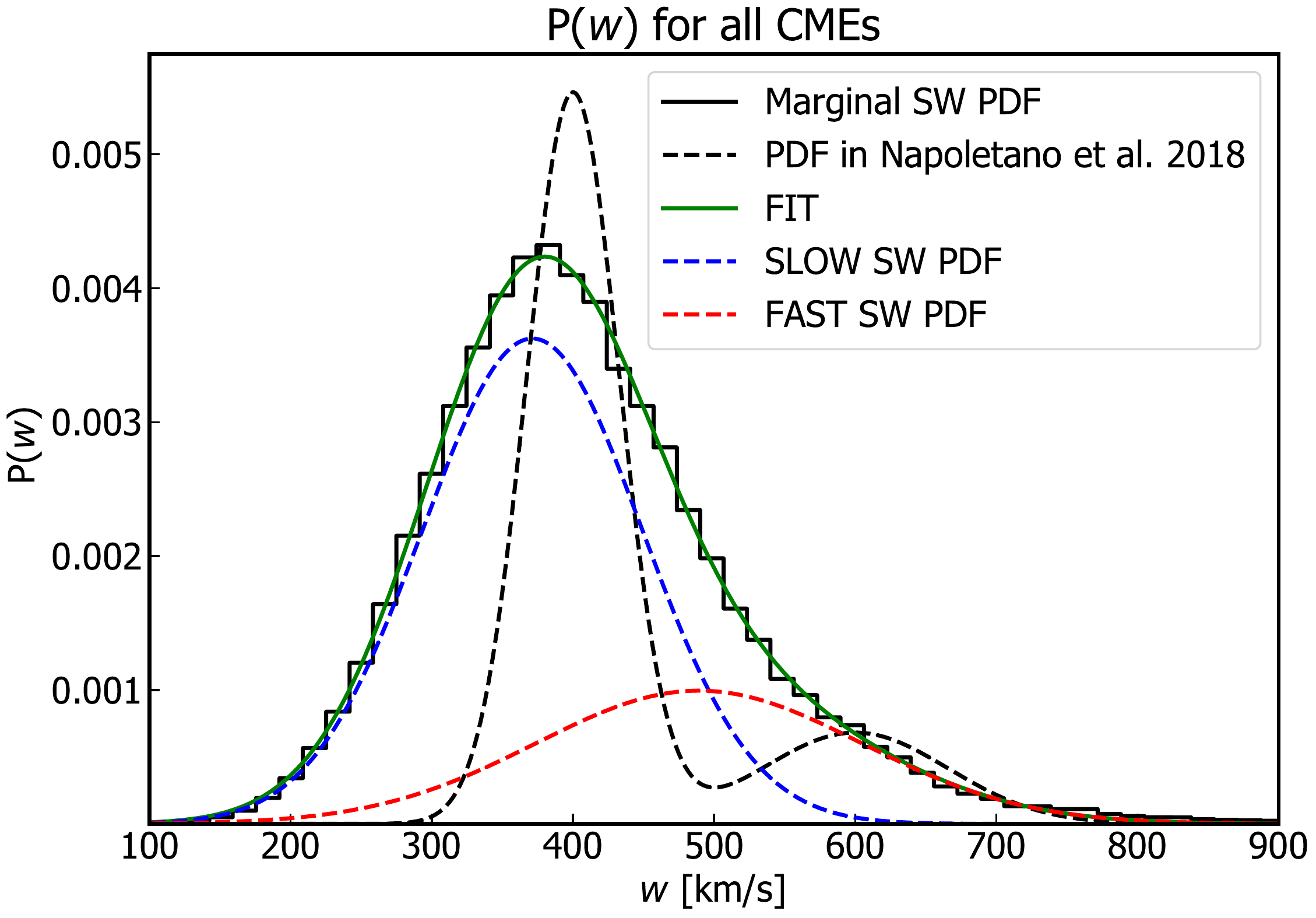}
\caption{The PDF of the solar wind speed from the whole dataset.  
The PDF was fitted with the sum of two Gaussian functions (green line), which can be interpreted as the slow solar wind (blue dashed line) and fast solar wind (red dashed line) contribution to the whole PDF($w$). As a reference, we also plot the SW PDF used in Paper~I (black dashed line)}.

\label{solar_wind_allevents}
\end{figure}
Similarly, we can extract the partial distribution of the solar wind values obtained by all the inversion procedures black line in Figure~\ref{solar_wind_allevents}).
We fitted this distribution with the sum of two Gaussian functions (blue and red lines), whose parameters are reported in Table~\ref{SW_table}.
\begin{table}[h!]
\centering
 \begin{tabular}{c c c c c}
 \hline
 \hline
 & $ A $ & $\overline{w} [\si{km/s}]$ & $\sigma_w [\si{km/s}]$ &   \\
 \hline
 blue Gaussian function & $3.6\cdot 10^{-3}$ & 370 & 80& slow solar wind\\
red Gaussian function & $1.0\cdot 10^{-3}$  & 490 & 100  & fast solar wind\\
 \hline
    \end{tabular}
    
    \caption{Parameters for the two Gaussian functions used to model the solar wind distribution.}\label{SW_table}
\end{table}
It is straightforward to interpret these two Gaussian functions in terms of slow and fastsolar wind.
We therefore redefine the parameters to generate the PDFs for the solar wind employed in Paper~I with the new values reported in Table~\ref{SW_table}.


\subsection{Testing PDBM performance with updated gamma and solar wind speed distributions}
\label{O_list_validation}
In order to test the new solar wind and drag parameter distributions and compare the performance of the PDBM related to old distributions (Paper I) our ICMEs list was used to compare the predictions with the observed values. The algorithm maps a large sample of initial conditions to the corresponding transit time and arrival speed through the DBM equations~(\ref{DBMr}) and~(\ref{DBMv}), allowing, in addition, to asses the forecast uncertainty from the output distributions. For each event, we start by randomly extracting an initial speed $v_0$ and a value for the solar wind speed $w$ from their respective normal distributions. If $v_0<w$ ($v_0>w$) the event is an accelerating (decelerating) CME, and a value for the drag parameter $\gamma$ is randomly generated from the corresponding log-normal distribution from subsection.~\ref{sub:PDFs}. Computed CME transit time $T_c$ and impact speed at 1AU $v_C$ are then obtained through the DBM equations with this set of initial conditions and parameters. For each event, a statistical distribution for ToA and SoA results from repeating this procedure a large number of times. The mean values $\langle T_c \rangle$ and $\langle v_c \rangle$ of such distributions, together with their standards deviations are taken as representative of the forecast for each event.

\begin{figure}[p!]
    \centering
    \includegraphics[width=15cm]{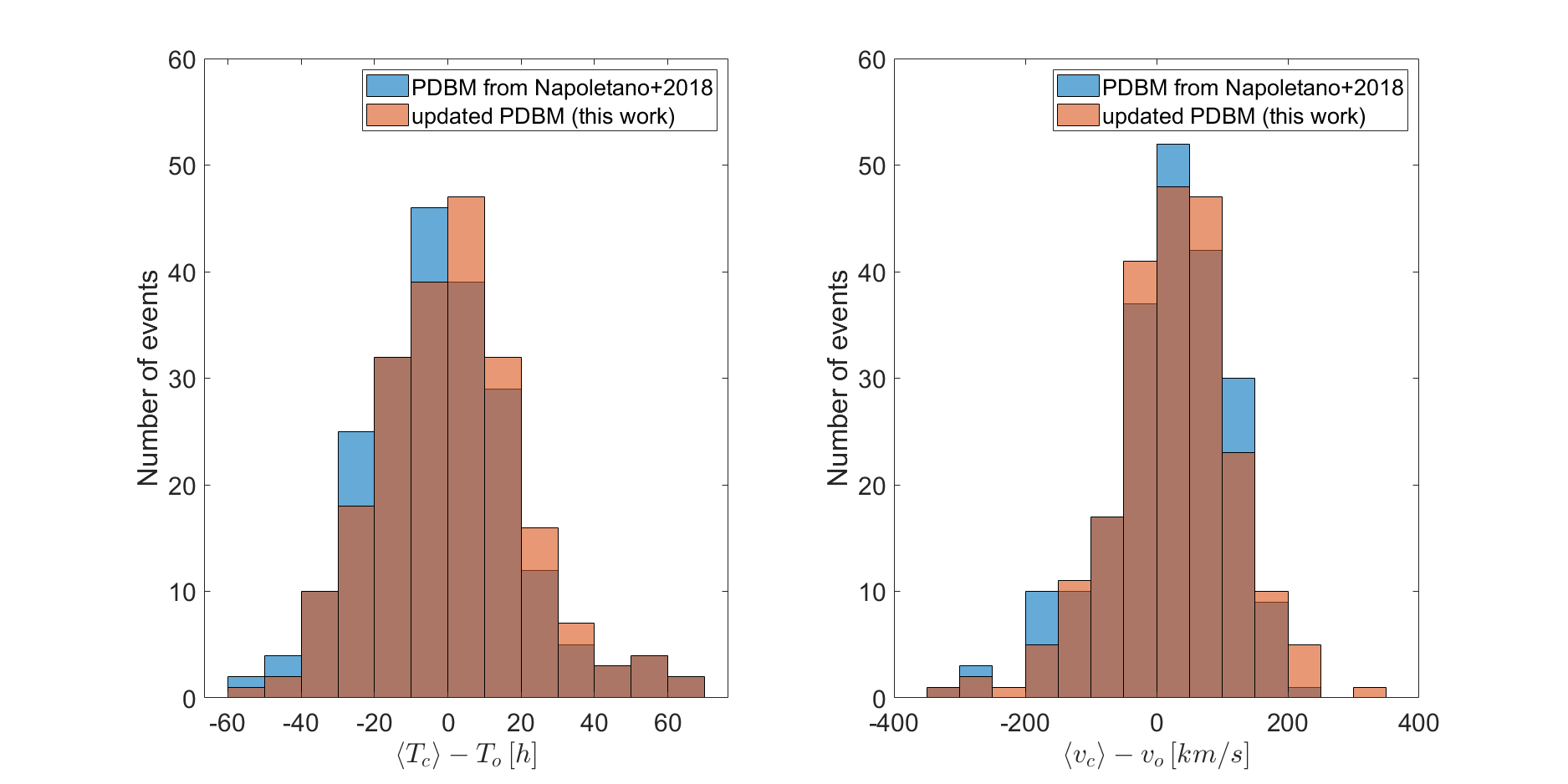}
    \caption{Histograms of the differences between computed and observed values for the ICME travel time (left) and arrival speed (right), computed using our ICME list. Such predicted values are computed both with the old solar wind and drag parameter distributions from Paper~I and the new ones.}
    \label{histograms_1}
\end{figure}

\begin{figure}[p!]
    \centering
    \includegraphics[width=15cm]{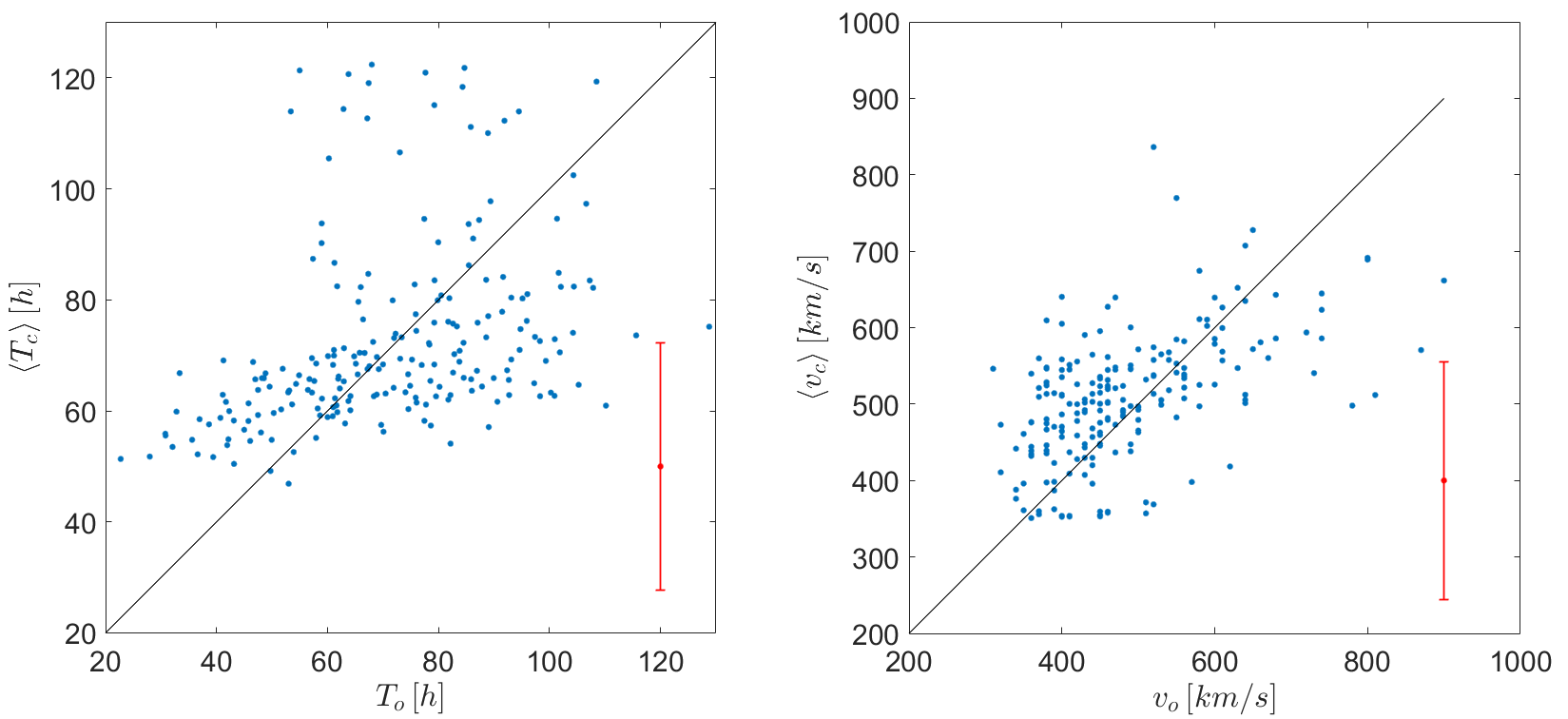}
    \caption{Plots of the computed mean ToA vs the observed ToA  (left) and of the computed mean SoA vs the observed SoA (right), obtained after applying the new PDBM distibutions on our dataset. Black line represents the perfect forecast. Due to the density of data points, vertical error bars have been omitted for sake of visualization, and the median error bar has been represented in the bottom right corner.}
    \label{plots_1}
\end{figure}

Figure~\ref{histograms_1} shows the histograms of the difference between transit time and arrival speed comparing the old PDBM and the new one on this set of events.
Relevant performance indicators are collected in Table~\ref{table_list_1}. Figure~\ref{plots_1} shows the plots of the PDBM prediction versus the observed values of the ICME transit time and impact speed. For 74\% of events, the observed ToA falls within the standard deviation of the predicted one, and for the arrival speed this occurs for 90\% of the events. This may be an indication of a possible overestimation of the error on the forecasted ToA, related to the errors on the input velocities.

\begin{table}
\centering
 \begin{tabular}{l c c c}
& & PDBM from paper I & New PDBM \\
 \hline
ME of arrival time prediction & $\langle \Delta T \rangle$  \text{[h]} & $-1.7 \pm 21.3$ & $1.1 \pm 20.6$\\
MAE of arrival time prediction & $\langle |\Delta T| \rangle$   \text{[h]}  & 16.4  & 15.9\\
RMSE of arrival time prediction  & RMSE($T$)\text{[h]} & 21.3 & 20.4\\
ME of arrival speed prediction & $\langle \Delta v_1 \rangle$  \text{[km/s]} & $14.2 \pm 105.7$ & $20.3 \pm 103.7$ \\
MAE of arrival speed prediction & $\langle |\Delta v_1| \rangle$ \text{[km/s]} & 79.1 & 78.0 \\
RMSE of arrival speed prediction & RMSE($v_1$)\text{[km/s]} & 106.4 & 111.7 \\
 \hline
    \end{tabular}
    \caption{Performance indicators for the application of the PDBM model from Paper~I and with the updated PDFs to the database presented in this work. 
    }\label{table_list_1}
\end{table}

\subsection{Validation against a subset of Paouris' ICME database}
\label{Pau_validation}
\begin{figure}[p!]
    \centering
    \includegraphics[width=15cm]{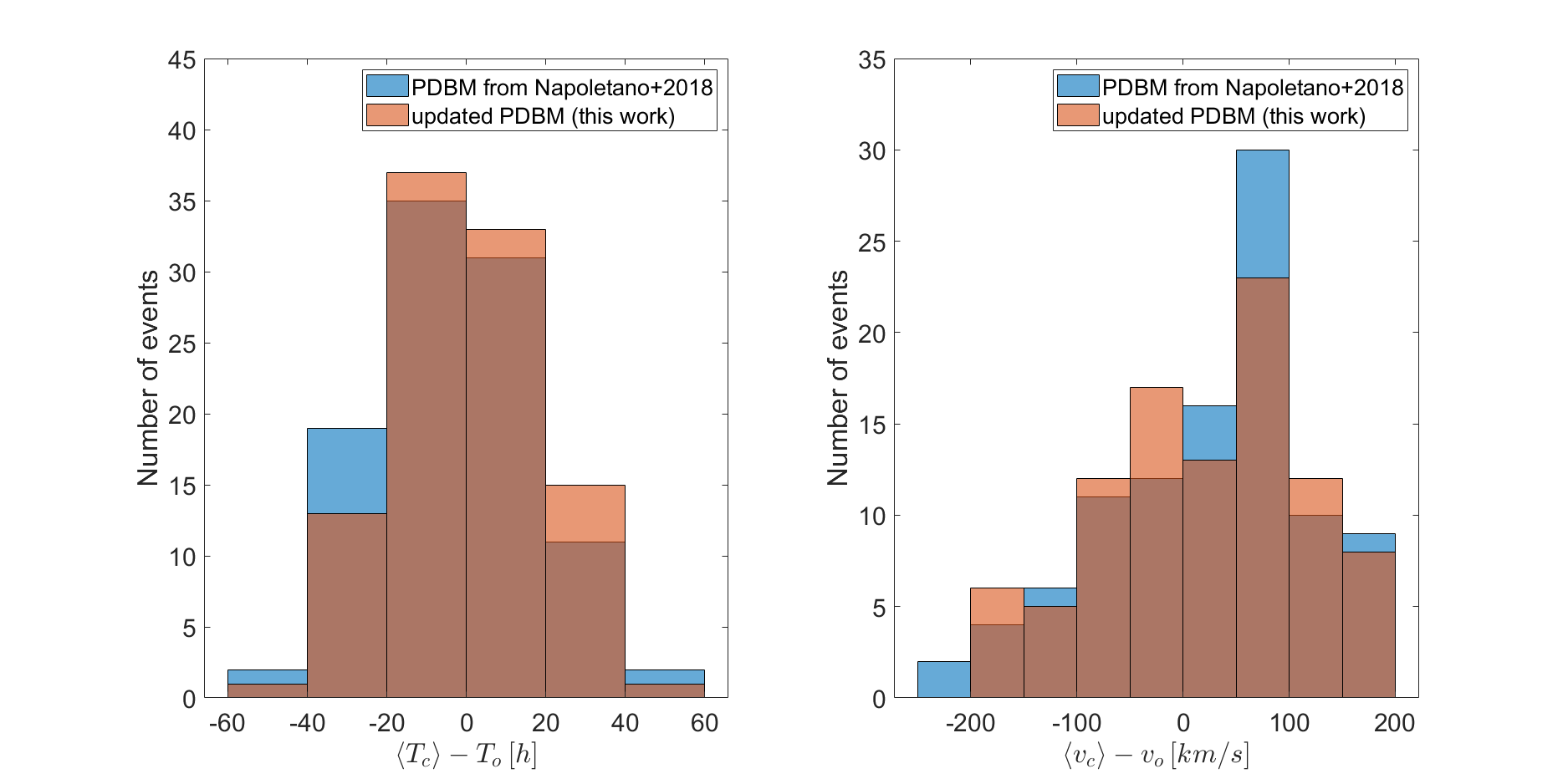}
    \caption{Histograms of the differences between computed and observed values for the ICME travel time (left) and arrival speed (right), computed on  on the subset of events from \cite{paouris2017interplanetary} list. Such predicted values are computed both with the old solar wind and drag parameter distributions from Paper~I and the new ones.}
    \label{histograms_2}
\end{figure}

\begin{figure}[p!]
    \centering
    \includegraphics[width=15cm]{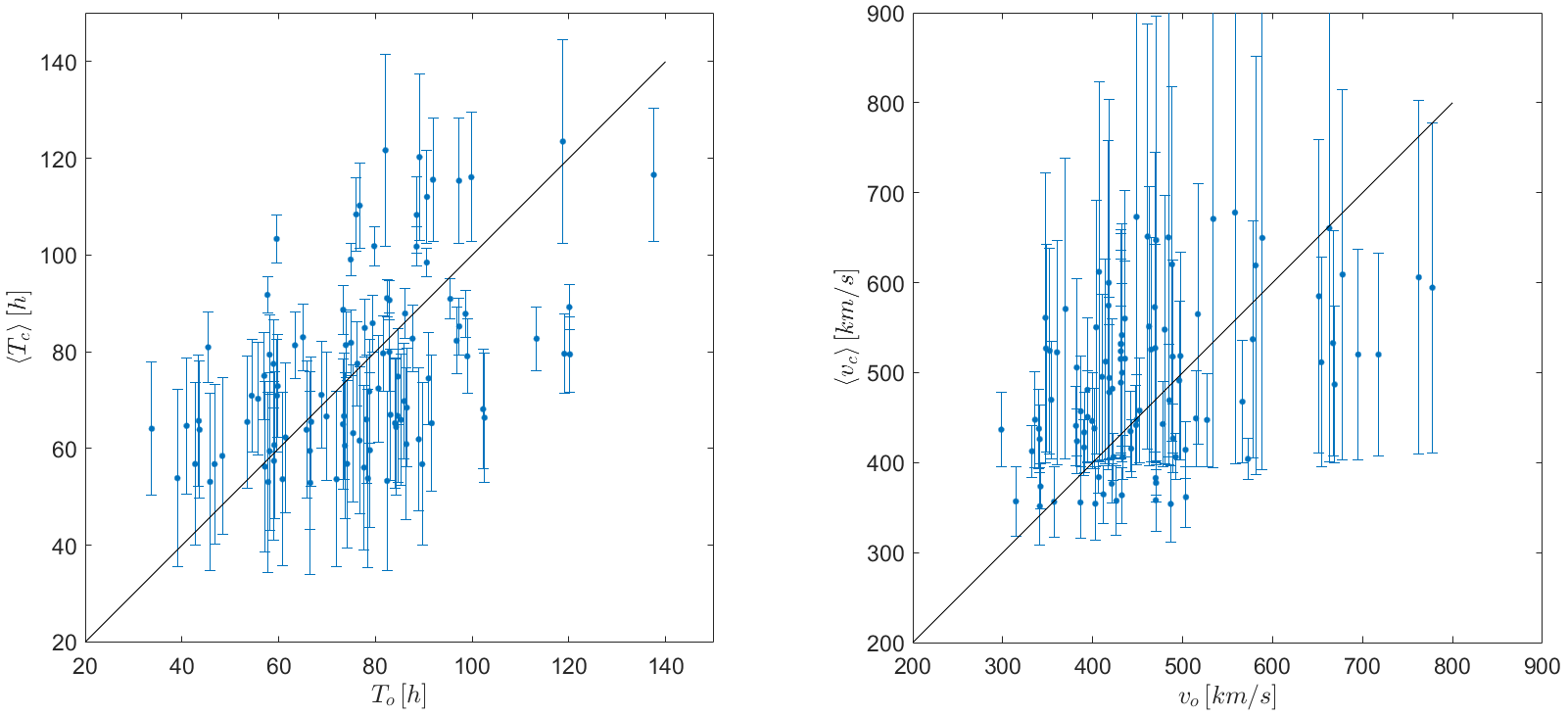}
    \caption{Plots of the computed mean ToA vs the observed ToA  (left) and of the computed mean SoA vs the observed SoA (right), obtained after applying the new PDBM distibutions on the subset of events from \cite{paouris2017interplanetary} list.}
    \label{plots_2}
\end{figure}

To test the validity of the new PDBM distributions on a list of past ICME events which is independent from the one described in Section~\ref{Data}, we employed a list of 100 events from the Hence we employed a list of 100 events from the \cite{paouris2017interplanetary} list, obtained after excluding 92 common ICMEs between the two databases.
Figure~\ref{histograms_2} shows the histograms of the difference between transit time and arrival speed comparing the old PDBM and the new one on this set of events.
Relevant performance indicators are collected in Table~\ref{table_list_2}.
Figure~\ref{plots_2} shows the plots of the PDBM prediction versus the observed values of the ICME transit time and impact speed. For 65\% of events, the observed ToA falls within the standard deviation of the predicted one, while for the arrival speed this occurs for 78\% of the events. 


\begin{table}
\centering
 \begin{tabular}{l c c c}
 & PDBM from paper I & New PDBM \\
 \hline
  ME of arrival time prediction & $\langle \Delta T \rangle$  \text{[h]} & $-3.0 \pm 19.9$ & $-0.2 \pm 19.5$\\
  MAE of arrival time prediction & $\langle |\Delta T| \rangle$   \text{[h]}  & 16.8  & 16.3\\
  ME of arrival speed prediction & $\langle \Delta v_1 \rangle$  \text{[km/s]} & $24.4 \pm 97.0$ & $30.0 \pm 102.4$ \\
  MAE of arrival speed prediction & $\langle |\Delta v_1| \rangle$ \text{[km/s]} & 84.1 & 88.8 \\
  RMSE of arrival time prediction & RMSE($T$)\text{[h]} & 14.4 & 15.5\\
 RMSE of arrival speed prediction &  RMSE($v_1$)\text{[km/s]} & 71.9 & 80.4 \\
 \hline
    \end{tabular}
    \caption{Performance indicators for the application of the PDBM model from Paper~I and with the updated PDFs to the database by \cite{paouris2017interplanetary}.
    }
\label{table_list_2}
\end{table}

\newpage
\section{Discussion and Conclusions}
\label{Disc+Conc}
We compiled a list of CME-ICME pairs with a reliable association between their remote observations and in situ signatures.
To evaluate the CME initial speed and launch time, we applied a polynomial fit procedure to coronagraph data.
We then employed an algorithm to establish the most suitable type of solar wind accompanying the CME.
We built this database by using the automated methods for CME characterization that are used for a CME detection and forecast service.
Therefore, the uncertainties reported in the database should be representative of the errors in  real-time use.\\
We employed this database to retrieve the parameters $\gamma$ and $w$ in the DBM equations.
In this procedure, we took into account the uncertainty on the observations, mapping the input PDFs into the PDFs for the model parameters.
The robust statistics, granted both by the larger ICME databasewith altogether 214 ICMEs and the Monte-Carlo-like inversion method, produced a joint PDF populated enough to allow us to verify the PDFs proposed in \cite{napoletano2018probabilistic}.

The empirical solar wind $w$ PDF has been modelled using the sum of two Gaussian functions, and given their parameters we interpreted these as representative of the slow and fast solar wind distributions.\\
Similarly, we verified that a Lognormal function fit is a suitable function for fitting the empirical PDF for the drag parameter $\gamma$.
Although the Lognormal is a long-tailed function, we can define the average value $\overline \gamma = 1.\times 10^{-7} \si{km^{-1}}$. 
Also, the fit parameters appear to be quite close to those of Paper I.
It is worth to note here that the  Drag Based Ensemble Model (DBEM) implementation by \citeA{dumbovic2018drag} uses as $\gamma$ PDF a Gaussian function with $\mu= 1.\times 10^{-8} \si{km^{-1}}$ and $\sigma= 0.5\times 10^{-8} \si{km^{-1}}$, that is ten times smaller than the value we found. Similarly to the present study, in a more recent paper \cite{calogovic2021}, they applied a reverse modeling procedure with the DBEM aimed to find optimal values for the DBM parameters. Their results showed that for the drag parameter a higher median value of about three times larger ($\gamma=0.32\times 10^{-7}\si{km^{-1}}$) and an extended range of values ($\sigma=0.7 \times 10^{-8} \si{km^{-1}}$) are needed than the one used in the previous version of the DBEM, for lower MAE and ME in predictions.
Interestingly, in the study of \cite{Rollett2016}, over the list of 21 ICME tracked by heliospheric imager, the average value of the fitted drag parameter is also generally larger than that of the previous models ($\approx 0.4\times10^{-7}\si{km^{-1}}$, even excluding 4 cases which yielded unrealistic too-high values for $\gamma$), still following the same trend. 
Furthermore, a recent paper by \cite{Paouris2021} found that the DBEM needs a larger value between 2.1 and 4.8 times larger) of the drag parameter to model a set of CMEs. 
%
%
%
The large statistics of the inverted DBM parameters, allowed us also to try and separate the PDFs for those CMEs which are accelerated or decelerated by the solar wind.
Although the statistics for the accelerated CMEs is much reduced, we found evidence that the $\gamma$ PDFs are significantly different.
In particular, the accelerated CMEs seem to experience, on average, a larger value of $\gamma$ and to equalize more rapidly their speed to that of the solar wind.\\
We therefore introduce refinement of the PDBM with respect to that presented in Paper~I, using different $\gamma$ PDFs in case of accelerated or decelerated ICME.
It is worth explaining why we chose to look for this separation. As the drag-based model describes the CME propagation taking as a reference the motion of a solid body in a fluid stream, a different value for the drag parameter may be expected if such body does not present the same shape to a fluid coming from the rear (accelerating CME) and to a fluid coming from the front (decelerated CME). We suggest that this may be the case, as ICMEs are typically depicted as curved flux tube, and our finding that the accelerated CMEs experience a higher drag than the decelerated ones is in accordance with such picture, where we expect a higher drag due to the fluid piling up in the rear of accelerating ICMEs, and a lower drag for decelerating ones, which undergo a smoother solar wind flow on their edges.
Interestingly, some results from \cite{vrsnak2008dynamics} may lead to similar conclusions. In this work, they investigated several relationships between the CME dynamics and the CME mass and found a correlation between the latter and the initial speed, concluding that since slower CMEs tend to have lower mass, larger values of $\gamma$ are expected, as the drag parameter is related to the inverse of the CME mass (refer to equations in \citeA{vrsnak2013propagation}).
%

The updated  method shows an improvement on predicted ToA average, both in the test against the initial database (sec.~\ref{O_list_validation}) and in the validation against Paouris' CME list (sec. ~\ref{Pau_validation}), although the performances obtained with the old and new PDBM implementations are comparable and well within the error bars.\\
%
%
It is worth to note that the performance of our model is in agreement with the results from the investigation of \cite{vourlidas2019predicting}, which investigated the relation between the size of the database and the ToA MAE for several forecasting methods. 
Apparently, ToA MAE in the range $10-15\si{h}$ is the current limit on the performance obtained by almost all the methods for ICME forecasting, including numerical models. 
Our interpretation is that this limit is set by both the lack of knowledge about the actual state of the interplanetary medium and the large uncertainties on the CME initial properties.\\

This triggers two considerations.
First, as long as the input data has such large errors, there will be probably little gain in tuning the model performance without taking into consideration those input errors.
This, of course, affects also the new PDBM implementation we are proposing in this work.
Second, it is important to test/validate the forecast procedures (especially those suited for real-time implementation) using a standard database of CMEs, in order to allow the comparison of the performance of models under the very same conditions.
To this purpose, we think that data sets such as that presented in this work and that in \cite{paouris2017interplanetary} will be of benefit to the CME modeling community when comparing the performances of different models and methods.\\
%
%
Lastly, it is worth to stress that the joint PDFs in Figure~\ref{image3} show a non-linear correlation between $\gamma$ and $w$: while we did not make use of this information in our work, the use of a joint PDF for the parameter extraction would reduce the parameter space by one degree of freedom.
This approach definitely deserves further investigation, and it will be treated in future work.
\section{Acknowledgements}
This work is in part supported by the ESCAPE project (the European Science Cluster of Astronomy $\&$ Particle Physics ESFRI Research Infrastructures) that has received funding from the European Union Horizon 2020 research and innovation program under the Grant Agreement n. 824064.
We also acknowledge support for our research by the project ``CEI6: Circumterrestrial Environment: Impact of Sun-Earth Interaction'' funded by the MIUR Progetti di Ricerca di Rilevante Interesse Nazionale (PRIN) Bando 2017 - grant 2017APKP7T.
EC is partially supported by NASA grant 80NSSC20K1580.
We made use of the Near-Earth Interplanetary Coronal Mass Ejections Since January 1996 catalog compiled by Ian Richardson and Hilary Cane and available at \url{http://www.srl.caltech.edu/ACE/ASC/DATA/level3/icmetable2.htm}
We made use of the CME catalog generated and maintained at the CDAW Data Center by NASA and The Catholic University of America in cooperation with the Naval Research Laboratory.
We made use fo the Heliophysics Event Knowledge database.
This research used version 0.9.8 of the SunPy open source software package \cite{sunpy_community2020, Mumford2020}.
SOHO is a project of international cooperation between ESA and NASA.
We acknowledge the two anonymous reviewers for the meticulous inspection of the manuscript and the insightful comments, which provided interesting ideas for extensions of the current work.
EC is partially supported by NASA grants 80NSSC20K1580  and 80NSSC20K1275. GN gratefully acknowledges the financial support from MIUR grant 2017APKP7T, under the supervision of Prof. Francesco Berrilli. The authors thank Prof. Francesco Berrilli for the valuable and profound discussion that lead to new ideas and a general revision of the manuscript.

\appendix
\section{Further details on Methods}\label{AppendixA}
Within the Appendix we provide more information about the methods employed to build the CME database.

\subsection{Identifying the CME source on the solar disk}\label{AppendixA1}
To find the most probable CME source on the Sun, we employed a source finding algorithm that makes use of HEK \cite<Heliophysics Event Knowledgebase ->[]{hurlburt2010heliophysics} to query which solar features (Active Regions, Solar Flares, Filaments Eruptions) that may have been the source of the CME, are within an area $A$ and a time span $\Delta t$ compatible with the CME launch parameters. 
The time span $\Delta t$ is defined by an estimate of the time and duration of the CME liftoff obtained from LASCO images. 
The search area $A$ is the whole solar sector defined by the CME POS angle and the angular width $W$ of the CME if $W <180^{\circ}$ (normal and partial halo CMEs).
In the case $180^{\circ} \leq W <270^{\circ}$ (half halo CME), the same sector is limited to $800"$ from the disk center.
In the case $W \geq 270^{\circ}$ (full halo CME), $A$ is the central part of the solar disk, within $600"$ from the disk center.
In the case multiple possible sources are retrieved by the query, the position of the CME source is the weighted average of the retrieved feature positions, with active regions weighting the larger between 0.1 and their associated $R$ values \cite{schrijver2007characteristic}, flares weighting 25 and filament ejections weighting 500.
In those cases where no potential feature is retrieved by the query, the default source position is computed as the intersection between the CME POS angle vector and a circle centred on the disk centre, with radius $R^*$, where $R^*=\frac{1}{3}R_\text{sun}$ in case of a full halo CME, $R^*=\frac{1}{2}R_\text{sun}$ in case of a half halo CME, $R^*=\frac{2}{3}R_\text{sun}$ in case of a partial halo CME, $R^*=R_\text{sun}$ in case the CME width is smaller than $90^{\circ}$.
To the CME foot-point position we associate an error given by the larger between $5"$ and the standard deviation of the weighted averaging described above.
We employed this method and these weighting values since they are those in use in the real-time CME detection and propagation services SWERTO \cite{berrilli2017swerto} and IPS \cite{veettil2019ionosphere} and have been set after an extensive test on a number of known CME-ICME counterparts.
\subsection{De-projecting the CME propagation vector}
\label{AppendixA2}
\begin{figure}[p!]
\centering
\includegraphics[width=1.0\textwidth]{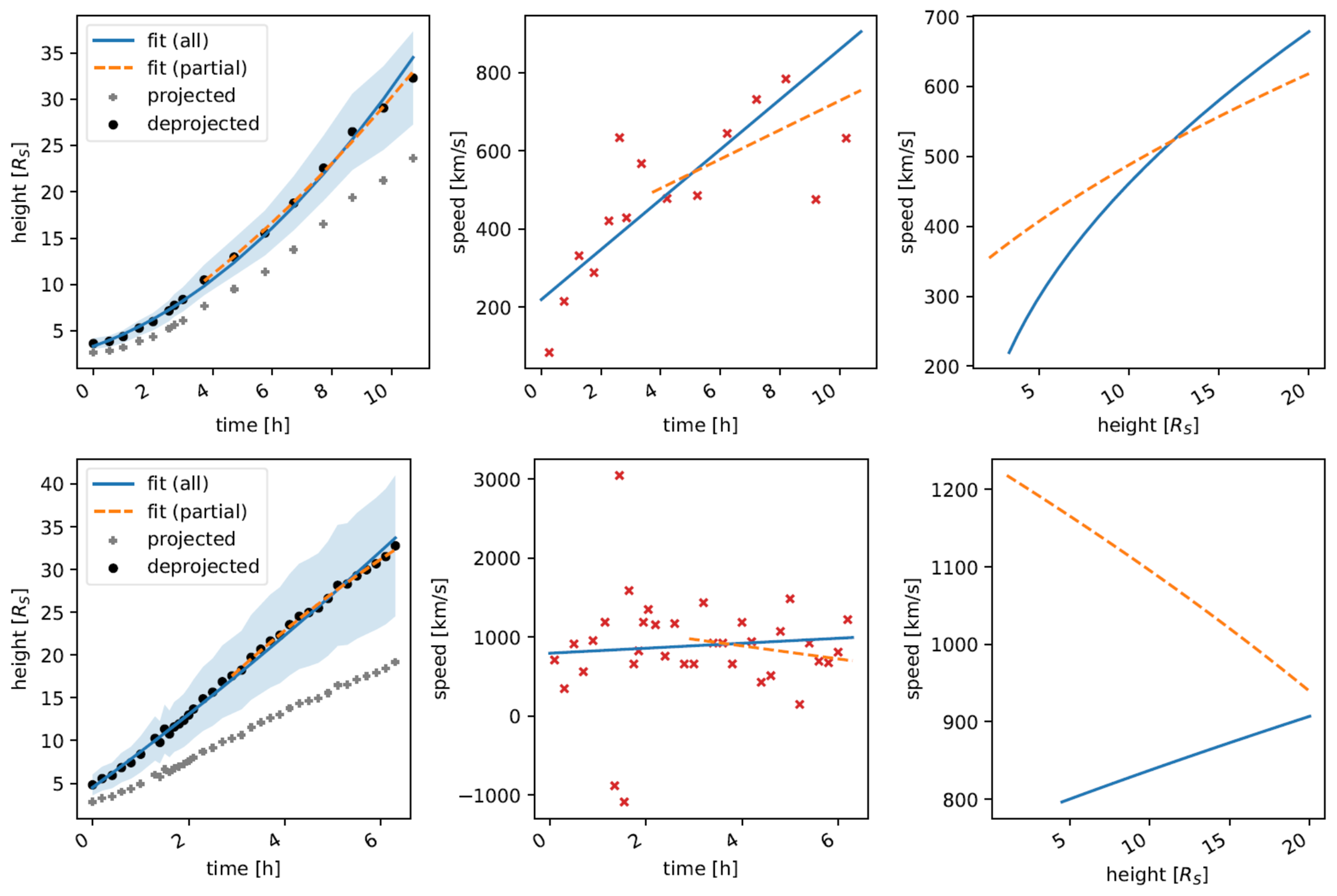}
\caption{Illustration of the fitting procedure used on the CME height versus time data. The two rows show examples for CME n.13 and n.179 from the dataset. Left column: the original and deprojected data, with the shaded area indicating deprojection uncertainties. Two types of quadratic fits are shown: using all the data points (used for our database for robustness) and using only part of the data. Middle column: CME speed as determined from the fits and as sampled from the deprojected data. Right column: CME speed versus CME height, as determined from both types of fits. At 20 solar radii, they usually agree quite well.}
\label{timedistanceplot}
\end{figure}

After the identification of CME most likely source region,it is necessary to compute the radial speed $v_r$ from the measured POS speed of the CME front. The procedure is based on equation 1 in \citeA{gopalswamy2010deprojection}, assuming a cone model for the CME shape (see Figure 1 in the same reference). The main difference is that we use the de-projection coefficient to obtain the de-projected position $R_r$ from the POS position of the CME front, instead of directly de-projecting the speed. By using such equation we can also compute the error associated to the radial position $dR_r$.\\
Once we have the data to plot a time-distance relation in the de-projected framework, we can obtain the radial speed by doing a linear or quadratic fit of the $r(t)$ relationship described by these data.
%
%
The standard option is to fit a quadratic relationship, but when the number of measured position of the CME available for the fit is less than 9, a linear fit is used. Examples of quadratic fits are presented in Figure~\ref{timedistanceplot}. Note that with a quadratic fit, the CME's acceleration is assumed to be constant and its speed is assumed to be a linear function. Because of this approximation, different velocities can be obtained depending on which part of the $r(t)$ data is used for the fit, as illustrated in the figure. In this paper, we have for robustness always used all available data points.
\\
With the parameters from the fit, we are able to compute several quantities of interest for the CME liftoff.
Namely, the time when the CME front reaches the $20 R_\text{Sun}$ distance and its associated error; the CME $v_r$ ($@ 20 R_\text{Sun}$) and associated error; the possible CME front residual acceleration at $20 R_\text{Sun}$ distance.
Those values make part of the CME liftoff characteristics in the database: start date, de-projected speed, de-projected speed error, acceleration. 
Since the error on the arrival date is negligible, the error on the Start Date is reported as transit time error. 

\subsection{Associating a solar wind speed to each CME}
\label{AppendixA3}

In order to propagate the ICME with an appropriate solar wind speed, for each event we have to hypothesize if the ICME interacted with a stream of slow (\textit{S}) or fast (\textit{F}) solar wind.
It is well known that coronal holes are sources of fast solar wind streams \cite{1973SoPh...29..505K, 1976SoPh...46..303N}, therefore we implemented an algorithm which discriminates the solar wind type by verifying if the CME source region is close to a coronal hole.
A suitable algorithm queries the HEK (Heliophysics Event Knowledge) catalog for all the Coronal Holes present on the solar disk. 

The time range queried starts from 4 hr before the estimated CME launch time to the CME launch time (considering the error).

As a consequence, we associate the event with fast (slow) solar wind if the CME source coordinates are close to (far from) any Coronal Hole retrieved by the query.

%

\section*{Data Availability Statement}

The ICME catalog built for the analysis in Section 3, together with a tool for the data visualization and the module employed for running the PDBM simulations, can be downloaded from \url{https://doi.org/10.5281/zenodo.5818470} \cite{Napoletano2021}. 

\bibliography{biblio}

\end{document}